\documentstyle[preprint,aps]{revtex}
\begin{document}
\draft
\title{
Spin-Charge Separation at Finite Temperature in the Supersymmetric {\it t-J}
Model with Long-Range Interactions
}
\author{Yoshio Kuramoto and Yusuke Kato}
\address{Department of Physics, Tohoku University,\\
Sendai 980-77, Japan}
\maketitle
\begin{abstract}
Thermodynamics is derived rigorously for the 1D supersymmetric {\it t-J} model
and its SU($K,1$) generalization with inverse-square exchange.
The system at low temperature is described in terms of spinons, antispinons,
holons and antiholons obeying fractional statistics.  They are all free and
make the spin susceptibility independent of electron density, and the charge
susceptibility independent of magnetization.
Thermal spin excitations responsible for the entropy of the SU($K,1$) model are
ascribed to free para-fermions of order $K-1$.\\

\end{abstract}


Elementary excitations in one-dimensional electrons consist of collective modes
of spin and charge with velocities different from each other.  This is often
referred to as the spin-charge separation.  In the supersymmetric {\it t-J}
model with the nearest-neighbor hopping, the spin velocity depends on the
average electron density $n$ per site \cite{bares}.
 In the model with long-range interactions, on the contrary, the spin velocity
is independent of $n$, and the charge velocity is independent of the
magnetization $m$ \cite{ky91,kawa92,wlc,haha92}.  We refer to this independence
as the strong spin-charge separation.
A question of basic importance is then whether the strong spin-charge
separation persists at finite temperature $T$.
The purpose of this paper is to derive thermodynamics of the model
microscopically.
We show that the low $T$ thermodynamics is determined by elementary excitations
obeying fractional statistics, but that the strong spin-charge separation
breaks down as the temperature increases toward a characteristic temperature.

At zero temperature elementary excitations in the long-range {\it t-J} model
have been investigated in \cite{haha94} where free semionic spinons, holons and
bosonic antiholons are identified.
The thermodynamics of the spin chain (Haldane-Shastry model) has been derived
with the use of the empirical supermultiplet rule \cite{hal91}.  The empirical
rule was generalized to the {\it t-J} model in writing down the basic equation
for the free energy \cite{wlc}.
The symmetry behind the supermultiplet is now identified as Yangian
\cite{haltani}.
On the other hand, a simple explanation for the supermultiplet follows by
deriving the family of long-range lattice models from the continuum Sutherland
model \cite{suth} in the limit of large coupling constant \cite{poly,ss}.  In
\cite{ss}, the asymptotic Bethe ansatz (ABA) was used to derive the spectrum of
the Sutherland model.
In this paper we also utilize the limiting procedure, but do not rely on any
unproved assumption.

The supersymmetric {\it t-J} model in one dimension \cite{ky91} is
represented in a form
\begin{equation}
{\cal H} = \sum _{i < j}\, t_{ij}(\tilde{P}_{ij}+1-2n_i),
\label{susy}
\end{equation}
where $n_i$ is the electron number operator at site $i$.
We have introduced a graded permutation operator
$
\tilde{P}_{ij} = \sum_{\alpha,\beta} X_i^{\alpha\beta} X_j^{\beta\alpha}
\theta_\beta
$
where $X_i^{\alpha\beta}$ changes the state $\alpha$ at site $i$ to $\beta$
with $\alpha, \beta$ being either 0 (vacant) or $\sigma = \pm 1$ (occupied by
either spin written also as $\uparrow, \downarrow$).  The sign factor $\theta
_\beta$ is $-1$ if $\beta=0$ and is 1 otherwise.
The interaction has the long-range form:
$ t_{ij}=t D(x_i-x_j)^{-2} $
with $D(x_i-x_j)=(L/\pi )\sin [\pi (x_i-x_j)/L]$ and $t >0$.  Here $L$ denotes
length of the system, and $x_i$ a lattice site.

We use as an auxiliary a variant of the Sutherland model in the continuum
one-dimensional space.
It is given by
\begin{equation}
{\cal H}_{Suth}=
         -{1\over 2m}\sum_{i=1}^N\frac{\partial ^2}{\partial x_i^2}
         +\frac{1}{4m}\sum_{i<j}
         \frac{\lambda(\lambda - M_{ij})}{D_{ij}^2 },\label{SP}
\end{equation}
where $M_{ij}$ is the exchange operator of coordinates of particles $i$ and $j$
\cite{poly}.
The spectrum of the model obtained by the ABA \cite{ss} has been proven to be
exact \cite{kaku}.
The ground state energy $E_0$ is of $O(\lambda ^2)$, and the energy $E$
relative to $E_0$ is given in terms of the distribution function $\nu_\alpha
(k)$ for the component $\alpha$ with momentum $k$.  Here $k$ is an integer
multiple of $2\pi/L$ with the periodic boundary condition.
The result is
\begin{equation}
E-E_0=\sum_k \frac{k^2}{2m} \nu \left(k\right)
                  +\frac{\pi\lambda}{4mL}\sum_{k,k'}\vert k-k'\vert
		   \nu \left( k \right)
		   \nu \left( k'\right),
\label{functional}
\end{equation}
where
$\nu (k) = \sum_\alpha \nu_\alpha (k)$.
For identical particles with internal degrees of freedom, the symmetry of the
wave function leads to $M_{ij}\tilde P_{ij} =-1$ within this Hilbert space.  We
take the limit of large $\lambda$ and $m$, keeping the ratio
$
t = \lambda/(m a^2)
$
fixed.  Here $a = L/N$ corresponds to the lattice constant.  In this limit the
$N$ particles crystallize at the lattice points,
and the first term without $\lambda$ in Eq.(\ref{functional}) becomes
negligible.  The second term describes the spectrum of the lattice model given
by Eq.(\ref{susy}) plus the  lattice vibration (phonon).
In the following we take the units such that $t = a = 1$.
It can be shown that the phonon frequency $\omega_q$ is given
for positive momentum $q$ by
$
\omega _q = q(\pi -q/2 ) = (\pi^2-p^2)/2 \equiv \omega (p),
$
where we have introduced $p=\pi -q$ and $\omega (p)$ for later convenience.

The quasi-particle energy $\epsilon (k)$ for the component $\alpha$ is given by
$
\epsilon (k) = \delta E/\delta\nu _\alpha (k)
$
which in fact is independent of $\alpha$.
The velocity (or rapidity) $p(k)$ is defined by
$ p(k) = \partial\epsilon(k)/\partial k $
which tends to $\pi$ as $k$ goes to infinity.
By further differentiating the rapidity we obtain
\begin{equation}
\frac{\partial p}{\partial k} =
\frac{\partial p}{\partial \epsilon} \frac{\partial \epsilon}{\partial k} =
\frac12 \frac{\partial p^2}{\partial \epsilon}
=\nu (k),
\label{p/k}
\end{equation}
in the thermodynamic limit $N \rightarrow \infty$.

The distribution functions are determined so as to minimize the thermodynamic
potential.  Namely one has
$
\nu_\sigma (k) = f\left(\epsilon(k)-\mu_{\sigma}\right)
$ and $
\nu_0(k) = b\left(\epsilon(k)-\mu_0\right)
$
where $f$ and $b$ are fermi and bose distribution functions with chemical
potentials $\mu_\alpha$.  Then integration of Eq.(\ref{p/k}) with respect to
$\epsilon$ gives
\begin{equation}
\frac12 (\pi^2-p^2 ) = \omega (p) = T\ln (1+\nu _0) -
T\sum_{\sigma}\ln (1-\nu_\sigma),
\label{pvs.e}
\end{equation}
where $\pi^2$ in the leftmost side comes from the boundary condition at
$\epsilon \rightarrow \infty$, and where obvious arguments of distribution
functions are omitted.  The results so far given are the same as those obtained
in \cite{ss} with use of the ABA, but are quoted here as prerequisite to the
new results to be given below.

It is convenient to introduce another distribution function $\rho_\alpha (p)$
in the $p$ space by
$ \rho_\alpha (p) dp = \nu_\alpha (k) dk $,
with the sum rule $\sum_\alpha \rho_\alpha (p) =1$ for each $p$.
Excitations in the {\it t-J} model can be described in terms of
$\rho_0 (p)$ for charge and $\rho_s(p) = \rho_\uparrow (p)-\rho_\downarrow (p)$
for spin.
At $T=0$, the distribution functions reduce to step functions: $\rho_0(p) =
\theta (p_c-p)$ and $\rho_s(p)=\theta (p-p_s)$ where
$
p_c = \pi (1-n)$ and $p_s = \pi (1-m).
$
Here $n = n_\uparrow +n_\downarrow$ is the average number of electrons per
site, and $m=n_\uparrow -n_\downarrow$ the magnetization.
Note that $p_c$ and $p_s$ corresponds to velocities of charge and spin,
respectively.

Magnetic and charge susceptibilities describe changes of $m$ and $n$ against
the changes of the magnetic field $h$ and the electron chemical potential
$\zeta$.  These are related to $\mu_\alpha$ by
$
\mu_\sigma -\mu_0 = \zeta + \sigma h.
$
The phonons are not affected by the change and can be disregarded. Let us
consider the low $T$ case where $P \equiv \exp(-\beta h) \ll 1$ and $M \equiv
\exp(-\beta \zeta) \ll P$. Then the thermodynamics is determined by excitations
near $p_c$ and $p_s$.
We shall first derive the charge susceptibility and introduce
\begin{equation}
\epsilon_c(p) = \frac12 (p^2-\pi^2)+2\zeta \equiv \frac12 (p^2-p_c^2),
\label{pvs.z}
\end{equation}
Then $\rho_0(p)$ near $p=p_c$ is obtained as
\begin{equation}
\rho_0 (p) = \frac{b(\epsilon -\mu_0)}{2+b(\epsilon -\mu_0)}
=\frac{1}{\sqrt {4\exp[\beta\epsilon_c(p)] +1}}
\end{equation}
where $f(\epsilon -\mu_\sigma)$ has been approximated by 1 with the condition
$M \ll P \ll 1$.
The function $2\rho_0(p)$ is the distribution function for semionic particles
with energy $\epsilon_c(p)$/2 \cite{berwu}.  This particle is called the holon.
 We shall discuss the statistics in detail later.

The density $n$ is given by
\begin{equation}
1-n = \int_0^\pi \frac{d p}{\pi} \rho_0 \simeq
\frac{1}{\pi}\int_{-\infty}^\infty d\epsilon_c p(\epsilon_c)\left(
-\frac{\partial \rho_0}{\partial \epsilon_c}\right),
\end{equation}
where we have used the delta-function like character of $-\partial
\rho_0/\partial \epsilon_c$ in extending the range of integration.  The
quantities
\begin{equation}
I_n \equiv \int_{-\infty}^{\infty} d \epsilon_c \epsilon_c ^n \left(
-\frac{\partial \rho_0}{\partial \epsilon_c}\right),
\end{equation}
are calculated to be: $I_0 =1,\ I_1 =0,\ I_2 = 2\pi^2 T^2/3, I_3 =
12\zeta(3)T^3$.  Thus one can perform a low $T$ expansion of $n$.
With the use of Eq.(\ref{pvs.z}) we obtain
$
p(\epsilon_c) \simeq p_c +\epsilon_c /p_c
-\epsilon_c^2/(2p_c^3)+\epsilon_c^3/(2p_c^5).
$
The charge susceptibility $\chi_c(n)$ is given by
\begin{equation}
\frac{\partial n}{\partial \zeta} \equiv \chi_c (n) = \frac{2}{\pi^2(1-n)}
[1+\frac{2T^2}{3\pi^2 (1-n)^4}] +O(T^3),
\label{charge}
\end{equation}
which is independent of $m$.
This independence is a signature of the strong spin-charge separation at low
$T$.
The presence of $O(T^3)$ term in Eq.(\ref{charge}) makes a difference from the
standard Sommerfeld expansion.
Analysis of the $p$ integral shows that the $m$-dependence at low $T$ enters
through an exponentially small parameter
$\exp (-T_{mix}/T)$ where
\begin{equation}
T_{mix} = \frac12 (p_s^2-p_c^2) = \frac12 \pi ^2 (2-n-m)(n-m).
\label{tsc}
\end{equation}

We now turn to the spin susceptibility. In deriving $\rho_s$ near $p_s$, we can
set $\rho_0=0$.  Then from Eq.(\ref{pvs.e}) we obtain with a little
manipulation
\begin{equation}
\rho_s (p) = \frac{1}{\sqrt{ 4\exp[\beta\epsilon_s(p)] +1}},
\end{equation}
where $\epsilon_s(p)$ is given by
\begin{equation}
\epsilon_s(p) = \frac12 (\pi^2-p^2)-2h = \frac12 ( p_s^2-p^2).
\label{pvs.h}
\end{equation}
Thus the spin excitation with energy $\epsilon_s(p)/2$ also obeys the semionic
statistics, and is called the spinon \cite{hal91}.  The magnetization $m$ is
given by integration of $\rho_s (p)$ and
the differential susceptibility is derived as
\begin{equation}
\frac{\partial m}{\partial h} \equiv \chi_m (m) = \frac{2}{\pi^2(1-m)}
[1+\frac{2T^2}{3\pi^2 (1-m)^4}]+O(T^3).
\label{mag}
\end{equation}
It should be emphasized that $\chi_m (m)$ has precisely the same functional
form as $\chi_c(n)$.
That $\chi _m(m)$ is independent of $n$ is another signature of the strong
spin-charge separation.
The spin susceptibility $\chi_s$ is related to the magnetic susceptibility
$\chi_m$ by
$\chi_m = 4\chi _s$.

If we take the limit of zero magnetic field first, i.e., $h/T \ll 1$, we have
$
\rho_s (p) = \beta h\exp \left[ \beta\omega(p)/2\right].
$
Then the susceptibility is given by
\begin{equation}
\chi_m = \frac{2}{\pi^2}\left( 1+\frac{2}{\pi ^2}T\right) +O(T^2),
\end{equation}
which has the $O(T)$ correction in contrast to Eq.(\ref{mag}).
The difference comes from the $p$-linear spinon spectrum near $p = \pi$.
For general temperature, $\chi_s =\chi_m/4$ at $h=0$ can be derived numerically
from
\begin{equation}
\chi_s = \frac{\beta}{4\pi}\int_0^\pi d p\frac{f(1-f)}{2f+b}.
\end{equation}
Figure 1 shows the results for various $n$.  It is clearly seen that the zero
temperature limit as well as the initial slope is independent of $n$.
This again demonstrates the strong spin-charge separation.  We remark that
there is no logarithmic singularity near $T=0$ in contrast to the Heisenberg
model \cite{tak}.  The absence confirms that the supersymmetric {\it t-J} model
is the fixed-point model for one-dimensional electrons
\cite{ky91,haha94,forre}.
As seen in Fig.1, $\chi_s$ does come to depend on $n$ with increasing $T$.  The
breakdown of the strong spin-charge separation already begins at $T$
substantially lower than $T_{mix}$ given by Eq.(\ref{tsc}).

Let us turn to the entropy $S_{tJ}$ per site of the {\it t-J} model which
consists of the boson part $S_0$, the fermion part $S_\sigma$ for each spin,
and minus of the phonon part $S_{ph}$. We first compute $S_0$ given by
\begin{equation}
S_0 = \int_0^\pi\frac{d p}{\pi} \frac{1}{\nu }[(b+1)\ln (b+1)-b\ln b ],
\end{equation}
where one may set $\nu = 2+b$ at low $T$ since the dominant contribution comes
from $p\simeq p_c$.  Eliminating $b$ in favor of $\rho_0$ we get by partial
integration
\begin{equation}
S_0 = \frac{\beta}{2\pi}\int_{-\infty}^\infty d \epsilon  p(\epsilon) \epsilon
\left( -\frac{\partial \rho _0}{\partial \epsilon}\right) \simeq \frac{\pi T}{3
p_c}.
\end{equation}
The fermion part can be derived for arbitrary magnetization.  We start with the
expression $
NS_\sigma = -\sum_k [\nu_\sigma \ln\nu_\sigma +(1-\nu_\sigma)\ln
(1-\nu_\sigma)],
$
and change the integration variable to $\epsilon(k)-\mu_\sigma$. At low $T$ the
dominant contribution comes from $\epsilon(k)-\mu_\sigma \simeq 0$.  Then we
get
\begin{equation}
S_\sigma = \frac{\beta}{\pi}\int_{-\infty}^{\infty} d \epsilon (\epsilon
-\mu_\sigma) k(\epsilon)\left( -\frac{\partial f(\epsilon
-\mu_\sigma)}{\partial \epsilon}\right) \simeq \frac{\pi T}{3p_\sigma},
\end{equation}
with $p_\sigma = \partial \epsilon (k)/\partial k$ at $\epsilon (k) =
\mu_\sigma$.
In the case of $m >0$, the contribution $S_\uparrow$ with $p_\uparrow = \pi$ is
the same as $S_{ph}$ and cancels each other, while $p_\downarrow = p_s = \pi
(1-m)$.  The final result is
\begin{equation}
S_{tJ} \simeq \frac{T}{3}\left( \frac{1}{1-n}+\frac{1}{1-m}\right) =
\frac{\pi^2}{6} T
\left(\chi_c+\chi_m\right),
\end{equation}
which also corresponds to the specific heat $\gamma T$.  The result describes a
two component (spin and charge) Tomonaga-Luttinger liquid, and proves previous
conjectures \cite{ky91,kawa92}.
In the opposite limit of high $T$, we may neglect the $p$-dependence of
$\rho_\alpha (p)$.  Then for fixed $n_\sigma$ we recover the obvious result
$
S_{tJ} = -\sum_\sigma n_\sigma \ln n_\sigma -(1-n)\ln (1-n).
$

The fractional statistics of excitations is better understood by generalizing
the supersymmetry to SU($K,1$).
If a particle obeys the exclusion statistics characterized by $g$, the
distribution function $\rho(p)$ obeys the equation
\begin{equation}
\rho^h(p)+g\rho  (p) =1,
\label{stat}
\end{equation}
where $\rho^h(p) = w \rho (p)$ is the distribution function of the
anti-particle (hole) \cite{berwu}.  The weight factor $w$ is equal to $1/\nu_0$
for the charge component, and to $(1-\nu_\alpha)/\nu_\alpha$ for the spin
component $\alpha = 1,\ldots, K$ in the SU($K,1$) model.
The particle-hole duality \cite{haha94,berwu} becomes apparent if one divides
both sides of Eq.(\ref{stat}) by $g\  (\neq 0) $ and makes a rescaling $dp
\rightarrow dp/g$.  Then it is seen that the anti-particle follows the $1/g$
statistics.  The energy of the anti-particle is $-1/g$ times that of the
particle with the same $p$.

The holon and antiholon distributions near $p = p_c$ are given by
$
\rho_0 = b/(K+b)
$ and $
\rho_0^h(p) = 1/(K+b)
$
at low $T$ with $M \ll P \ll 1$.
Because of the relation
$ \rho_0(p)+K\rho_0^h  (p) =1 $,
the antiholon obeys the $K$ statistics without the rescaling of $dp$.  The
energy of the antiholon is given by $-\epsilon_c(p)$.  The holon on the other
hand obeys the $1/K$ statistics and $p$ needs the rescaling $dp \rightarrow
dp/K$.  Thus the compensating factor $K=2$ has appeared as $2\rho_0(p)$ for
holons in the SU(2,1) {\it t-J} model.

Similarly $\rho_\alpha$ near $p=p_\alpha$ for $\alpha = 1,\ldots, K$ is
rewritten as
$
\rho_\alpha = \nu _\alpha/(\alpha -1+ \nu_\alpha )
$
where the occupation $n_1 >  n_2 > \ldots > n_K > 0$ has been assumed.
Then we get
\begin{equation}
\alpha \rho_\alpha (p) +(\alpha -1)\rho_\alpha^h(p) = 1
\label{staspi}
\end{equation}
where $\rho_\alpha^h(p)$ is the distribution function of the anti-particle
(spinon).  Dividing  Eq.(\ref{staspi}) by $\alpha$ we obtain the statistics of
the spinon as
$(\alpha -1)/\alpha$ with the rescaling $dp \rightarrow dp/\alpha$.  The
anti-particle of the spinon (antispinon) correspondingly follows the $\alpha
/(\alpha -1)$ statistics with the rescaling $dp \rightarrow dp/(\alpha -1)$.
The antispinon is first identified in this paper.  It has not been noticed in
\cite{haha94} probably because the singlet ground state with $p_\alpha =\pi$
for all $\alpha$ has no antispinon.
In the special case of $\alpha = 2$ the statistics of spinon is reduced to
$1/2$, i.e. to the semionic one.
The rescaling explains why the spinon spectrum \cite{haltani} is periodic in
$\pi$ instead of $2\pi$.

For thermal excitations a description different from the exclusion statistics
can be more convenient.
As is well known the entropy of the {\it X-Y} chain can be understood most
easily in terms of free fermions introduced by the Jordan-Wigner
transformation.
For spin components more than 2, ref.\cite{poly94} has introduced para-fermions
for another long-range model with harmonic confinement potential.
We now explore this type of description in the present model.
For simplicity we consider the case of $n=1$ without magnetic field. Setting
$b=0$ in Eq.(\ref{pvs.e}) we obtain
$
\nu_\alpha = 1-\exp(-\beta\omega/K)
$
with $\omega = (\pi^2-p^2)/2$
for $\alpha = 1,\ldots,K$.  Then the fermion part of the entropy becomes
\begin{equation}
\sum_{\alpha =1}^K S_\alpha = \int_0^\pi \frac{d p}{\pi}[\frac{\beta\omega
/K}{\exp(-\beta\omega/K)-1}-\ln(1-\exp(-\beta\omega/K)].
\end{equation}
The phonon part $S_{ph}$ is given by the same expression as above but with
$K=1$.  Therefore, the entropy $S_{SU(K)}$ of the long-range model is written
as
$S_{SU(K)} = -\partial \Omega_{SU(K)}/\partial T $
with
\begin{equation}
\Omega_{SU(K)}
= -T\int_0^\pi \frac{d p}{\pi}\ln
[1+e^{-\beta\omega/K}+e^{-2\beta\omega/K}+\ldots +e^{-(K-1)\beta\omega/K}].
\label{para}
\end{equation}
This is in fact valid for any $T$.  It is natural to interpret $\Omega_{SU(K)}$
as that of ideal para-fermions of order $K$ for which up to $K-1$ particles can
take the same quantum number $p$.
The para-fermion is reduced to the fermion in the case of $K=2$, which can also
be regarded as a pair of spinons \cite{ft}.
In the case of $n < 1$, the holon also contributes to the entropy and
Eq.(\ref{para}) describes the spin part at $T \ll T_{mix}$.

In summary we have derived thermodynamics of the supersymmetric {\it t-J} model
and have shown that the system at low $T$ is equivalent to a set of ideal
particles  obeying fractional statistics.  The strong spin-charge separation is
caused by the absence of interaction among them.
We note that description of fractional statistics is the same as that of
single-component systems only at low $T$.
At higher $T$ the multi-component character of fractional statistics appears
explicitly.
Detailed discussion for general $T$ with extensive numerical results will be
given separately.
The authors thank N. Kawakami for useful discussions.

\noindent
Figure 1.
The spin susceptibility against temperature for various fillings. The unit of
energy is $t$.


\begin{references}
\bibitem{bares} P.-A. Bares, G. Blatter and M. Ogata, Phys. Rev. B{\bf 44}, 130
(1991).
\bibitem{ky91}Y. Kuramoto and H. Yokoyama, Phys. Rev. Lett. {\bf 67}, 1338
(1991); H. Yokoyama and Y. Kuramoto, J. Phys. Soc. Jpn {\bf 61}, 3046 (1992).
\bibitem{kawa92}N. Kawakami, Phys. Rev. B {\bf 46}, 3191 (1992).
\bibitem{wlc} D.F. Wang, J.T. Liu and P. Coleman, Phys. Rev. B {\bf 46}, 6639
(1992).
\bibitem{haha92}Z.N.C. Ha and F.D.M. Haldane, Phys. Rev. B {\bf 46}, 9359
(1992).
\bibitem{haha94}Z.N.C. Ha and F.D.M. Haldane, Phys. Rev. Lett. {\bf 73}, 2887
(1994).
\bibitem{hal91}F.D.M. Haldane Phys. Rev. Lett. {\bf 67}, 1338 (1991).
\bibitem{haltani}F. D. M. Haldane, in {\it Correlation Effects in
Low-Dimensional Electron Systems} (edited by A. Okiji and N. Kawakami,
Springer-Verlag, 1994) p.3.
\bibitem{poly}A. P. Polychronakos, Phys. Rev. Lett. {\bf 69}, 703 (1992).
\bibitem{ss}B. Sutherland and B. S. Shastry, Phys. Rev. Lett. {\bf 71}, 5
(1993).
\bibitem{suth}B. Sutherland, Phys. Rev. A {\bf 4}, 2019 (1971); {\it ibid.}
{\bf 5}, 1372 (1972).
\bibitem{kaku}Y. Kato and Y. Kuramoto, Phys. Rev. Lett. {\bf 74}, 1222 (1995).
\bibitem{berwu}Y.-S. Wu, Phys. Rev. Lett. {\bf 73}, 922 (1994).
\bibitem{tak}S. Eggert, I. Affleck and M. Takahashi, Phys. Rev. Lett. {\bf 73},
332 (1994).
\bibitem{forre}P.J. Forrester, Phys. Lett. A{\bf 196}, 353 (1994).
\bibitem{poly94}A. Polychronakos, Nucl. Phys. B{\bf 419}, 553 (1994).
\bibitem{ft}L.D. Faddeev and L.A. Takhtajan, Phys. Lett. A{\bf 85}, 375 (1981).

\end{references}
\end{document}